# WEB SERVICE SELECTION BASED ON RANKING OF QOS USING ASSOCIATIVE CLASSIFICATION

Molood Makhlughian[1], Seyyed Mohsen Hashemi[2], Yousef Rastegari[3] and Emad Pejman[1]

[1]Department of Computer Engineering, Islamic Azad University-South Tehran Branch, Tehran, Iran

{molood.makhlughian, emad.pejman}@gmail.com

[2]Dean of Software Engineering and Artificial Intelligence Department, Islamic Azad University-Science and Research Branch, Tehran, Iran

hashemi@isrup.com

[3]Department of Electrical and Computer Engineering, Shahid Beheshti University, Tehran, Iran

y_rastegari@sbu.ac.ir

## ABSTRACT

*With the explosive growth of the number of services published over the Internet, it is difficult to select satisfactory web services among the candidate web services which provide similar functionalities. Quality of Service (QoS) is considered as the most important non-functional criterion for service selection. But this criterion is no longer considered as the only criterion to rank web services, satisfying user's preferences. The similarity measure (outputs–inputs similarity) between concepts based on ontology in an interconnected network of semantic Web services involved in a composition can be used as a distinguishing criterion to estimate the semantic quality of selected services for the composite service. Coupling the semantic similarity as the functional aspect and quality of services allows us to further constrain and select services for the valid composite services.*

*In this paper, we present an overall service selection and ranking framework which firstly classify candidate web services to different QoS levels respect to user's QoS requirements and preferences with an Associative Classification algorithm and then rank the most qualified candidate services based on their functional quality through semantic matching. The experimental results show that proposed framework can satisfy service requesters' non-functional requirements.*

## 1. INTRODUCTION

Service Oriented Computing (SOC) and its predominant incarnation as Web Services have emerged as a powerful concept for building software systems [1].

Web service technology offers a potential solution for developing distributed business processes and applications, which can be accessible via the Internet. Considering the rapid increase of Web users and the growing complexity of their demands, simple atomic services are inadequate. When individual Web services are not able to meet complex requirements, SOC provides a flexible framework for reusing and composing existing web services in order to build value-added composite services. At first the functionalities required for these complex requirements (namely the tasks) and their interactions, the control and data flow, are identified. Secondly, an appropriate implementation must be selected and bounded to each task. However with the rapidly growing number of available services, a large number of services can be found for realizing every task which can provide expected functionality for each of them. So it leads to the issue of selecting the best Web services among a list of "Candidate Web services", with the same functionalities. These services are, of course, different from one another in non-functional properties such as response time, availability, throughput, security, reliability, and execution cost [2], and are therefore different in terms of efficiency.

A specific issue emerges to this regard is selecting the best set of services based on their Quality of services (QoS) to participate in the composition, meeting QoS constraints set by users. QoS is a measure for how well a service serves the customer.

Dynamic service environments cause some difficulties in service selection. As the services' availability cannot be known a priori, or QoS conditions fluctuate in such environments, service selection and composition must be performed at runtime. Therefore the execution time of service selection algorithms is heavily constrained, whereas the computational complexity of the problem is NP-hard [3]. Hence, finding an optimal composite service may not be practical. Due to changing conditions in such environments there is no guarantee that the selected services for composite service will be available at runtime or its QoS will not be fixed concerning the advertised one in WSDL.

In this paper, we present a service selection algorithm that copes with above issues. This algorithm consists of two phases. In the first phase, we use a classification data mining algorithm to classify web service candidates into different QoS levels respect to the defined QoS constraints form the user and using the result of this classification to define a utility value for each of the service candidates. In the second phase we focus on composing the best services of each task and more specifically on their functional level that aims to selecting and inter-connecting web services by means of their semantic connections.

The remainder of this paper is structured as follows. In the next Section we give an overview of related works. In Section 3, we present our service selection approach and give the details of it, and we conduct a set of experiments to evaluate its timeliness and optimality in Section 4. Finally, in Section 5, we conclude with a summary of our contributions and the future perspectives of this work.

## 2. RELATED WORKS

The approaches aim at determining the optimal service composition using brute-force-like algorithms (i.e., the execution time and costs are exponential even if simplifications are used), new approaches are satisfied with finding a nearly-optimal solution. These approaches are using (meta-) heuristics. (Meta-) heuristics are general search methods that exclude a huge number of solutions, because they do not consider optimal solution in their population. Hence (meta-) heuristics are more efficient than exact algorithms. The main disadvantage however is that they do not find the optimal solution in most cases, because, the exclusion of solutions is based on assumptions, thus there is no guarantee, that the excluded solutions do not contain the optimal one.

All existing approaches can be divided into exact approaches finding the optimal service composition and (meta-) heuristics selecting a nearly service composition. Instead of exact algorithms, (meta-) heuristics seem to be the better choice for QoS-aware service selection and composition.

Beside many proprietary heuristics [4,5,6], currently the Genetic Algorithm is the most promising heuristic in QoS-aware service selection and composition. C. Jaeger et al. [7] present a heuristic based on the genetic algorithm. A big challenge related to the Genetic Algorithms is the choice of the genetic operators: selection, mutation and recombination, which have a big influence of the efficiency and correctness of the algorithm. As these parameters are chosen randomly, the order in which service candidates are checked is chosen randomly. On the other hand, as the genetic algorithm can run endlessly, the users have to define a constant number of iterations fixed a priori, and fixing a high number of iterations does not give any guarantee about the quality of the result. Therefore, the genetic algorithm is deemed not useful for the purpose of selecting near-optimal compositions.

In recent years, some approaches (e.g., [8,9]) use the power of Data mining algorithms in knowledge extraction and pattern discovery among the huge amounts of data in web service selection.

Wu et al. [8] present a Bayesian network based Qos assessment model for web services. That could predict the service capability in various combinations of users' Qos requirements. This approach is used to evaluate the capability of each service, and the one with best capability is selected as the binding service. Though it uses Bayesian network classification algorithm for each provider/service to predict the level of QoS, it is computationally complex and is based on probabilities, moreover it just considers local constraints in web service selection and doesn't mention the global constraints.

Ben Mabrouk et al. [9] present a heuristic approach for service composition in dynamic environments. This solution uses the K-means algorithm to classify the web services to QoS levels, then it uses the result of this clustering in an utility function in order to rank web service candidates for each task as a local selection part, then it uses a search tree to select the best services to form the composition plans in an ordered way. The proposed solution is computationally expensive in both of the clustering algorithm and the structure of the search tree in a composition plans with the high number of activities, also it suffers from the deficiency of the clustering techniques, because clustering is an example of unsupervised learning. Furthermore it does not mention the semantic matching between output and input of the services.

## 3. QOS-AWARE SERVICE SELECTION ALGORITHM

Our proposed approach starts form the assumption that the user uses a graphical user interface to define his/her request. With the help of this interface user can express his request in terms of QoS requirements and his relative preferences for each of them. Then the interface formulates it as a machine-understandable specification. The composition plan is given from the expert that its functional tasks and all the controls and data flow between them are characterized. For every activity in the composition plan, a service discovery phase gives a set of service candidates able to fulfill the functional aspect.

The proposed service selection method consists of a heuristic algorithm based on classification data mining algorithm, CBA [10] algorithm. This classification allows for classifying candidate web services with respect to the QoS requirements and preferences defined by the user into a set of different QoS levels. Further with considering the functional aspect we use these levels to determine the utility of each candidate service to do a near-optimal selection.

As the outlined process of the proposed method is depicted in figure 1, our heuristic approach deals with the service selection problem through the following phases:

1. Scaling phase, which is a pre-processing phase to normalizing QoS values.
2. Service selection by local classification, which classifies candidate services according to different QoS levels and determines the utility of each candidate service.
3. Ranking based on functional aspect, which uses the results of the last phase to selecting the best services according to the functional aspect.

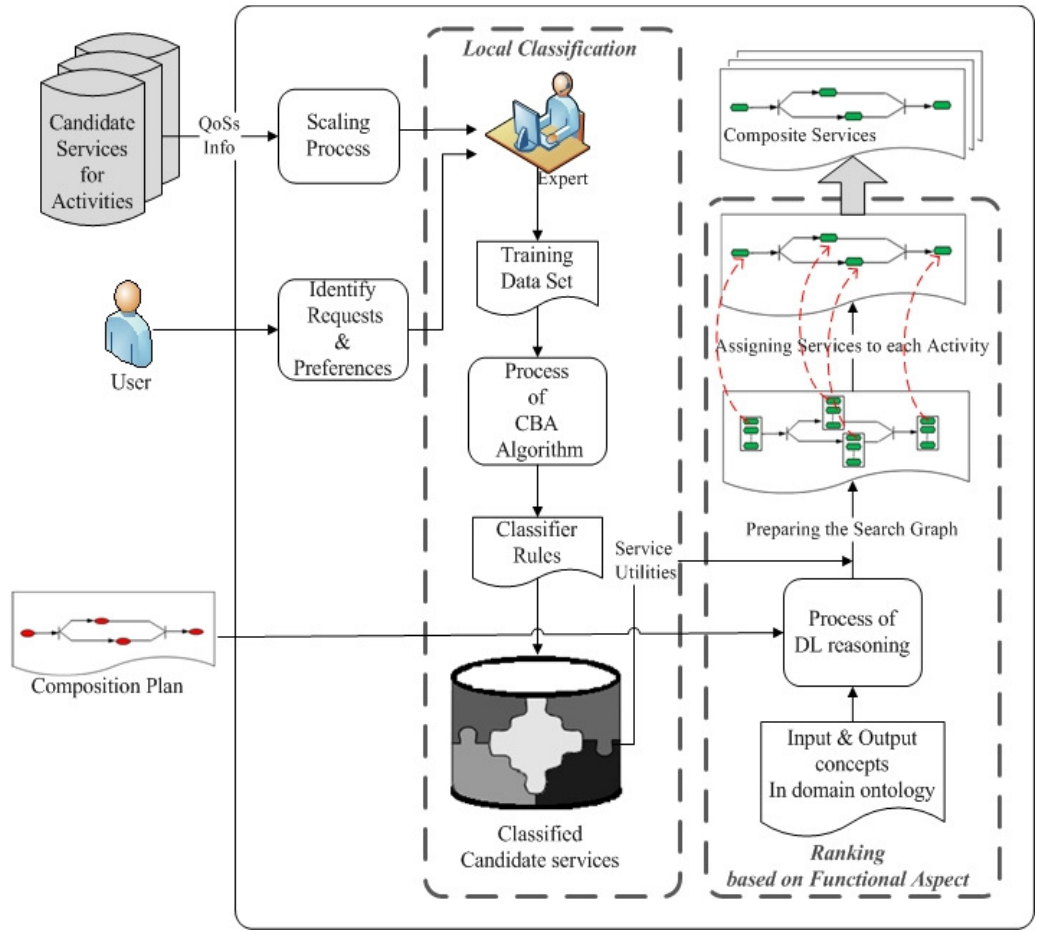

Figure 1. The outlined process of the proposed method

## 3.1. Scaling Phase

QoS attributes could be either positive or negative, thus some QoS values need to be maximized, (i.e., the higher the value, the higher the quality) for example availability and reliability, whereas other values have to be minimized, (i.e., the higher the value, the lower the quality). This includes criteria such as execution time and execution price. To cope with this issue, the scaling phase normalizes every QoS attribute value according to the following formulas [11]. The values of negative attributes are normalized by expression (1) and the values of positive attributes are normalized by (2).

$$Q'_{i,j} = \begin{cases} \frac{Q_j^{max} - Q_{i,j}}{Q_j^{max} - Q_j^{min}}, & if\ Q_j^{max} - Q_j^{min} \neq 0 \\ 1, & else \end{cases} \quad (1)$$

$$Q'_{i,j} = \begin{cases} \frac{Q_{i,j} - Q_j^{min}}{Q_j^{max} - Q_j^{min}}, & if\ Q_j^{max} - Q_j^{min} \neq 0 \\ 1, & else \end{cases} \quad (2)$$

Where $Q'_{i,j}$ signifies the normalized value of QoS attribute j associated with candidate service $S_j$. It is computed using the current value $Q_{i,j}$ and $Q_j^{max}$ and $Q_j^{min}$ which respectively denote the maximum and minimum values of QoS attribute j among all the candidate services.

After the data normalization is completed, all the QoS attributes values are lying in the [0, 1] interval and also have the same utility increase direction.

## 3.2. Service Selection by Local Classification

The classification is performed on candidate web services for all the activities in the abstract service composition. The purpose of this phase is to classify candidate web services to the multiple QoS levels with respect to the user's preferences on QoS attributes. In this classification each level consists of a set of candidate services that have approximately the same QoS value. These levels determine the relative importance of candidate services for selecting near-optimal compositions. To perform this classification we use an Associative Classification technique, CBA algorithm.

### 3.2.1. Classification Overview

CBA algorithm is based on Associative Classification which is the integration of two important data mining techniques, Classification rule mining and association rule mining. Associative Classification is a novel strategy performing classification where model of classification is based on a set of association rules in which consequent of each rule is restricted to contain only class attribute values.

Some definitions of associative classification are introduced here:

Definition 1 (Training Data Set $D$). If an object can be described by features $A_1, A_2, A_3, ., A_m$ and each object belongs to some class in a finite set of classes $C$, then training data set $D$ is a collection of instances $< v_1, v_2, v_3, ., v_m, c >$ where $v_i$ ranges over the domain of feature vector $A_i$ and $c$ represents the class of the object. As a preprocessing step any continuous valued attribute need be discretized.

Definition 2 (Future Object $O$). An instance $<v_1, v_2, v_3, ., v_m>$ whose class label is not known will be called future object and classifier will be used to predict its class.

Definition 3 (Class Association Rules: CARs). CARs will be mined from Training data set $D$. A special subset of association rules in which the precedent of rule contains the form of attribute value pairs and the consequent is restricted to take only class attribute value. Like association rules, CARs are also subjected to a minimum support and confidence threshold values. For instance a rule $r$, $A_i\ v_1, A_j\ v_2, \ldots ! c_k$ is a CAR and supp($r$), conf($r$) respectively denote support and confidence of rule.

CBA algorithm is an ordered rule algorithm based on coverage analysis. It consists of two parts Rule Generator and Classifier Builder. The goal of rule generator part is to mine all CARs from $D$ that their support and confidence are above minimum threshold values. Rule generation approach is based on Apriori [12]. Classifier builder uses class association rules (CARs). The goal is to short list high confidence rules from CARs to form a classifier. CARs are sorted in the descending order of their precedence. We call this set of CARs $R$. To build the classifier we select high precedence rules $r_i$ in $R$ to cover $D$. Finally classifier takes the form $<r_1, r_2, r_3, ., r_m, default\_class>$. $default\_class$ is the most probable class for all items. If all rules in the classifier are unable to classify a future object then it will be classified according to default class.

### 3.2.2. Defining QoS Levels

In order to classify candidate services, first we need to determine the QoS levels. To this regard we define $n$ QoS levels, where the value of $n$ is defined with respect to the service density and the expected accuracy for producing service classes.

Further the training data set is defined by the expert and the number of classes is the number of QoS levels and the attributes are requested QoS attributes defined by the user.

Each class defines the attributes which their values have a defined distance from the range of the values that specified from the user's request. (i.e., the first class shows the services which the value of their QoS attributes are in the same range with the user's demand and the second class shows the services which the value of their QoS attributes are one level lower than the user's demand and so on).

### 3.2.3. Computing the Service Utilities

The objective of our proposed algorithm is selecting the best services for the near-optimal compositions. Indeed, having a large number of choices for services during dynamic binding increases the number of alternative service compositions and subsequently a large number of compositions prevent the starvation in dynamic service environments.

In this regard, we define a utility function $U_i$ which shows the relative importance of a candidate service $S_i$. This function is computed based on two parameters: the first one is the QoS level which the service is classified in it through the classification, it represents the importance of the service (i.e., if it belongs to a class that its attributes are closer to the user demands, it will be better choice for selecting). The second one is the overall quality of the service. Concerning to this parameter, the service with the higher quality has the higher ability to be selected.
This function is computed as follows:

$$U_i = Class_i * Ave_i \qquad \text{Where} \qquad Ave_i = \frac{\sum_{j=1}^{n} Q'_{i,j}}{n} \qquad (3)$$

Where $Class_i$ is the coefficient of the class to which service $S_i$ belongs. $Ave_i$ is the total quality of service $S_i$, which computed as the average of the normalized QoS attributes values.

## 3.3. Ranking Based on Functional Aspect

The aim of this phase is selecting the best services for the near-optimal compositions such that 1) consider the best quality regard to the user's demand and 2) the composition of these services has high functionality.

### 3.3.1. Considering the Most Eligible Services

In order to select services for the near-optimal compositions we use the utility function that we have computed from the result of the classification part. So we do not consider all the possible combinations of services. In this regard, we just mention the candidate services that have a utility value over a defined threshold. This threshold allows us to focus on the most qualified services to our regard.

For defining threshold we consider two aspects: the number of required compositions and the execution time of the algorithm. Indeed, if we decrease the value of this threshold, the number of considered candidate services decreases. Consequently, the number of compositions and the execution time of the algorithm decreases respectively. Tuning the trade-off between these two aspects will make our algorithm adaptable; hence it could be applied to multiple dynamic service environments according to their constraints.

### 3.3.2. Functional Aspect of the Composition

Unlike most of approaches which just focus on the quality of composition by means of non functional parameters (i.e. QoSs), the quality of semantic links can be considered as a distinguishing functional criterion for semantic web service compositions.

Here we focus on the functional level of composing the candidate web services. The functional criteria of semantic link, was introduced for the first time in [13] which defined as a semantic connection between an output of a service and an input parameter of another service. Since the qualities of these connections are valued by a semantic matching between their parameters, semantic link composition could be estimated and ranked as well. Through the results of these estimations some compositions are inappropriate. Indeed a composite service which does not provide acceptable quality of semantic links might be useless as a service that does not provide the desired functionality. Indeed the semantic connection between Web services is considered as essential to form new value-added Web services.

Here we address the problem of optimizing in service selection with respect to this functional criterion. In other words, we focus on the aspects of selecting a set of appropriate service candidates for each task. We define an objective function In order to consider this aspect, preferences and constraints which are defined by end-user.
#### 3.3.2.1. Semantic Links

In semantic web, input and output parameters of web services are concepts referred to an ontology $T$, where the OWL-S profile [14], SA-WSDL [15] or WSMO capability [16] can be used to describe them (through semantic annotations).

At functional level, web service composition consists of retrieving some semantic links between output parameters $Out\_s_i \in T$ of service $s_i$ and input parameters $In\_s_j \in T$ of other service $s_j$. Such a link i.e., semantic link, $sl_{i,j}$ (Figure 2) between two functional parameters of $s_i$ and $s_j$ is formalized as $\langle s_i, Sim_T(Out\_s_i, In\_s_j), s_j \rangle$. Thereby $s_i$ and $s_j$ are partially linked according to a matching function $Sim_T$. This function expresses which matching type is employed to chain services.

The range of $Sim_T$ is reduced to the four well known matching types:

- **Exact** If the output parameter $Out\_s_i$ of $s_i$ and the input parameter $In\_s_j$ of $s_j$ are equivalent; formally, $T \vDash Out\_s_i \equiv In\_s_j$.

- **PlugIn** If $Out\_s_i$ is sub-concept of $In\_s_j$; formally, $T \vDash Out\_s_i \sqsubseteq In\_s_j$.

- **Subsume** If $Out\_s_i$ is super-concept of $In\_s_j$; formally, $T \vDash In\_s_j \sqsubseteq Out\_s_i$.

- **Intersection** If the intersection of $Out\_s_i$ and $In\_s_j$ is satisfiable; formally, $T \nvDash Out\_s_i \sqcap In\_s_j \sqsubseteq \bot$.

- **Disjoint** Otherwise $Out\_s_i$ and $In\_s_j$ are incompatible i.e., $T \vDash Out\_s_i \sqcap In\_s_j \sqsubseteq \bot$.

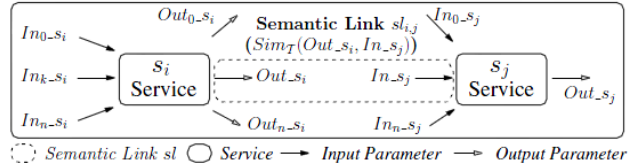

Figure 2. A Semantic Link $sl_{i,j}$ between $s_i$, $s_j$.

### 3.3.2.2. Semantic Link Quality

Several candidate services are grouped together in every task of an abstract composition. A method to differentiate their semantic links is to consider their different functional quality criteria. In this way, we use the semantic link quality model introduced in [17].

Here We consider the matching quality as the quality criteria for the semantic links $sl_{i,j}$ defined by $\langle s_i, Sim_T(Out\_s_i, In\_s_j), s_j \rangle$. The Matching Quality $q_m$ of a link $sl_{i,j}$ is a value in $(0, 1]$ defined by $Sim_T(Out\_s_i, In\_s_j)$ i.e., 1 for Exact matching type, $\frac{3}{4}$ for PlugIn, $\frac{1}{2}$ for Subsume and $\frac{1}{4}$ for Intersection.

The Disjoint match type is not considered since $Out\_s_i \sqcap In\_s_j$ is satisfied.

Given the above quality criteria, the quality vector of a semantic link $sl_{i,j}$ is defined as follows:

$$q(sl_{i,j}) = \big(q_m(sl_{i,j})\big) \quad (4)$$

In case of services $s_i$ and $s_j$ related by more than one semantic link, the value of each criterion is retrieved by computing their average.

**3.3.2.3. Generation of the Search Graph**

Our algorithm uses the structure of a graph for selecting the most appropriate candidate services for the composition. Moreover in this graph we use a priority queue to exploring the candidate services in a defined order corresponding to their importance of QoSs and semantic similarities with other services. These priority queues reduce the time spend for building and traversing the search graph by inclusion the unexplored candidate services in themselves. In addition they can increase the speed of search to acquiring the optimal composite services according to their importance, by visiting the first services of each queue. Hence this structure will decrease the computational complexity of traversing the graph and memory usage in the cases which the abstract service compositions comprises a large number of tasks or there are many candidate services for each task.

This graph is built from the candidate services according to the following rules:

- Each node represent a task in the abstract service composition;
- The order of these nodes is the corresponding order in the composition plan (i.e., if there is a link from task $T_x$ to $T_y$ in the abstract service composition, then the corresponding node of $T_y$ will be after the corresponding node of $T_x$);
- These nodes are made of a priority queue which consists of all the service candidates for the corresponding task which its $U_i$ is above the defined threshold as we have introduced before;

Candidate services for each task go to the priority queue in this order:

- For the tasks which do not have any incoming task in the composition plan, web services stay in the queue based on their utility values, i.e. the service with the highest utility value ($U_i$) stays at the top of the queue;
- For the other tasks, web services stay in the queue based on the following formula:

$$F_i = U_i * q(sl_{i,j}) \qquad (5)$$

Where $F_i$ is the final utility value for each candidate service based on their QoS values that respects the user needs and the matching quality of their semantic links between them and their preceding service;

To build the search graph, first of all we add the nodes correspond to the tasks which do not have any incoming task. Then we select a web service which is at the top of its queue. To select the services for the other nodes, we compute the matching quality of semantic links between the selected services and all the candidate services for the subsequent nodes then based on these values and the utility values (which was computed in the local selection part), stay the candidate services for these nodes in the priority queue based on their $F_i$ values. As the previous nodes, the service which is top of the queue is selected as the chosen service. And so on the other services for each node will have selected which the candidate services of each node stay in their corresponding queue based on the value of their $F_i$.

The selected services from top of the queues form the near-optimal composite service. This composite service is the first service that have the highest importance and unexplored services in each queue are the candidates for alternative composite services which when the services in the first composite service are no longer available or their QoS decreases (e.g., due to network disconnection or weak network connectivity) during the execution of the composition, could be replace with it.

The introduced structure of the search graph, allow for producing all the alternative composite services by selecting the unexplored candidate services in each queue. But on the other side, producing all the alternative composite services needs more computation so it is a time consuming process (for calculating semantic similarities between services), also saving these composite services need high memory, hence for tuning the trade-off between memory usage and the execution time of the algorithm, we produce only the first alternative composite service which is the best alternative.

We also propose a method for the case when a service which takes part in the composite service is no longer available, we could propose another composite service by replacing the unavailable service with another candidate service in its corresponding node of the search graph. In this case, for switching to another composite service we must rearrange the priority queue in that corresponding node, based on new utility values. This utility value is as the same as the $F_i$ function but the quality of the semantic links will be calculated as the follow:

At first we calculate the semantic quality of the connection between the selected service in the previous node with the services for this node and the semantic quality of the connection between the selected service in the following node with the services for this node, then the new semantic quality value for the new $F_i$ is computed based on the average of these two computed quality for each service in this node.

## 4. EXPERIMENTAL EVALUATION

### 4.1. Experimental Settings

We conducted experiments on an Intel(R) Core(TM) Duo CPU, 2.53GHz, 4GB RAM and a windows 7 operating system. In this experiment we focus on the execution time of our method. This metric measures the response time of our algorithm with respect to the size of the problem in terms of the number of activities and the number of services per activity. In these experiments, we measure the execution time of the ranking phase.

For the quantity allocation of the service qualitative parameters we use the QWS real dataset [18, 19]. This dataset includes measurements of 9 QoS attributes for 2500 real web services. The dataset was measured using commercial benchmark tools for web services, which were located using public sources on the Web, including UDDI registries, search engines and service portals.
In these studies, the authors provide a set of QoS metrics (i.e., response time, throughput, availability, validation accuracy, reliability). We use these metrics as a sample input data for our algorithm.

To accomplish the classification, we used the CBA[1] tool for implementing the classification algorithm. This tool has a graphical user interface and is designed particularly for implementing the CBA algorithm. Hence for this part we do not measure the response time.
For classifying the candidate services we have defined 3 different quality levels. These levels are specified to the distance of the QoS attributes to the user demands.
After the classification carried out, the classes' coefficient were defined as follow: 1 for services in the first class, $\frac{3}{4}$ for services in the second class and $\frac{1}{4}$ for services in the third class.

In general, on Contrary to QoS given by providers, the quality of semantic links are estimated according to DL reasoning. Matching quality of each semantic link has been inferred in a pre-processing step of semantic reasoning. Standard DL reasoning inference is achieved by means of a DL reasoner Fact++[2] [20].

---

[1] http://www.comp.nus.edu.sg/~dm2/

For the purpose of these experiments, we vary the number of activities and the number of services per activity between 10 and 50. The number of QoS constraints is fixed to 4 constraints (i.e., Response time, Availability, Throughput and Reliability) and for the sake of precision each experiment is executed 20 times and finally we calculate the mean value of the obtained results for our evaluation.

## 4.2. Performance Results

In figure 3 we demonstrate the execution time of ranking phase for the calculated near-optimal composite service. These measurements are obtained by fixing the number of QoS constraints to 4 and varying the number of activities and the number of service candidates per activity between 10 and 50. The obtained measurements show that the execution time of our algorithm increases along with the number of activities and the number of services per activity, which is an expected result.

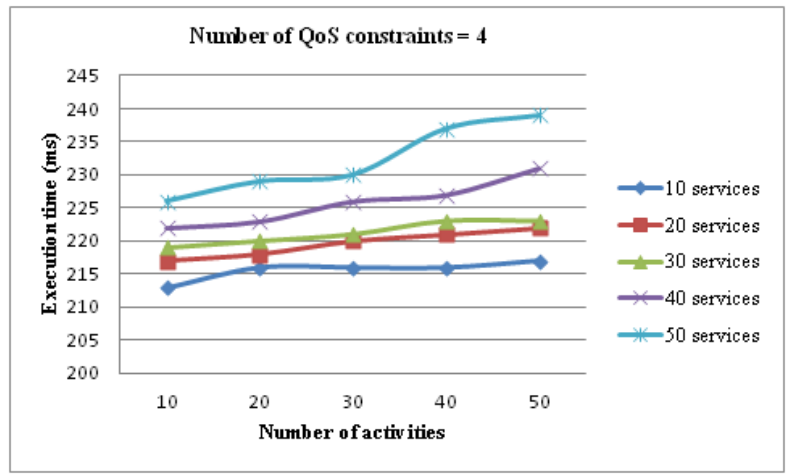

Figure 3. Execution time of the ranking phase for near-optimal composite service

Also figure 4 demonstrates the execution time of ranking phase for the first alternative composite service. The increment of the execution time compared with the figure 3 is due to the increment of the number of tasks and candidate services so the spending time for calculating semantic similarity of output-input concepts increases subsequently. DL reasoning is the most time consuming process in large-scale problem of quality-driven semantic web service composition (i.e., number of tasks and candidate services greater than 100 and 350).This is caused by the large number of potential semantic links between tasks.

---

[2] http://owl.man.ac.uk/factplusplus/

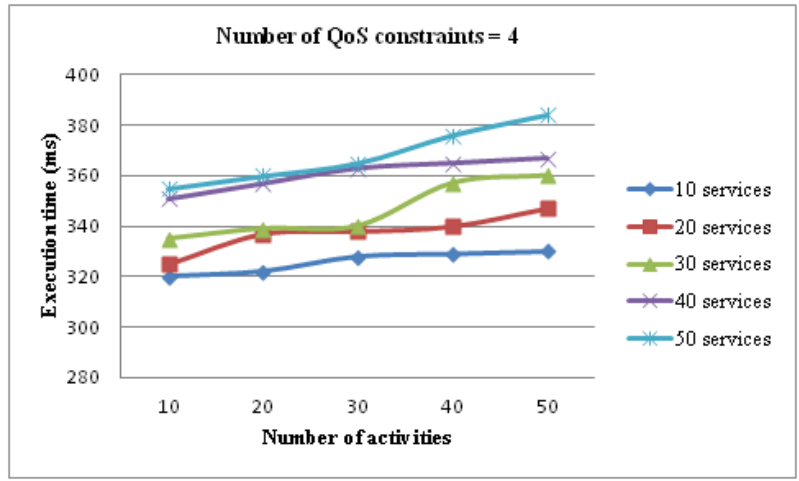

Figure 4. Execution time of the ranking phase for first alternative composite service

As the figure 3 and 4 show, our method has a good execution time compared with the evolutionary algorithms which the later algorithms (e.g., genetic algorithms, PSO) take a long time to execute composite service (e.g., for a number of activities more than 25, it takes 2847 and 2236 ms, respectively for the PSO and GA algorithms [21]).

Moreover unlike our proposed method which aims to produce the composite services in an ordered way, the evolutionary algorithms produce the composite services in an unordered way and there is no evidence that the first composite service which produced by these algorithms be the optimal one, in other words the next produced composite services may be better than the first one.

## 5. CONCLUSION

As the number of the services published over the internet is growing at a very fast pace, selecting satisfactory web services among the candidate web services which provide the same functionalities is difficult. QoS is considered as the most important non-functional criterion for further filtering services. Besides this criterion, the quality of semantic links can be considered as a distinguishing functional criterion for semantic web service selection and composition.
We address the web service selection problem by defining constraints within a quality model to balance QoS metric with functional quality. The functional quality evaluates the quality of semantic links between the semantic description of output-input parameters of web services, while QoS focuses on the non-functional criterion to retrieve the satisfactory services regard to the user's requirements and preferences.

Our objective has been to address service selection in the context of a QoS-aware middleware for dynamic service environments. To do so we have proposed an approach which at first by using the CBA algorithm, aims to classify the candidate web services to different QoS levels, differentiate the services within each class, respect to their distances from the user's demand for the QoS criterion. By this classification a utility value is defined for each service that shows its relative importance. In the next part, by the use of the structure of a graph, we have ranked the candidate services by defining the final utility of each service considering the results of the classification phase and their functional quality through measuring the semantic similarity between their output-input concepts. Finally by producing alternative composite services in an ordered way, we could cope with changing conditions of dynamic service environments.

Our proposed approach has four advantages: First, it uses the classification data mining algorithm to specify the most eligible services respect to the user demand. Applying data mining algorithms to this field brings new ideas. Second, by producing alternative composite services satisfying QoS constraints, our algorithm could cope with changing conditions in dynamic service environments. Third it shows a satisfying capability in terms of execution time, which it is an important point in dynamic service environments and finally with applying the semantic similarity between services by semantic links it increases the accuracy of selection.

Our proposed method makes part of our ongoing research by using strong data mining algorithm in order to decrease the execution time and improve the optimality of our heuristic algorithm, and besides the local optimization, by combining the global constraint of the user to the local constraints, considering the QoS constraints through the whole composition to ensure meeting global QoS constraints.

**Authors**

**M. Makhlughian** received her B.Sc. degree in Software Engineering, from the Azad University-Parand Branch, Tehran, Iran in 2009 and her M.Sc. degree in Software Engineering from the Azad University-South Tehran Branch, Tehran, Iran in 2012. Her research interests include Web Service technology, Service-Oriented Computing, Service Annotation, Selection and Composition, Ontology Engineering and Data Mining.
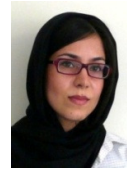

**Dr. S. M. Hashemi** received his M.S. degree in Computer Science from Amirkabir University of Technology (Tehran Polytechnic University) in 2003, and his PhD degree in Computer Science from the Azad University in 2009. Moreover, he is currently a faculty member at Science and Research Branch, Azad University, Tehran. His current research interests include Software Intensive Systems, E-X systems (E-Commerce, E-Government, E-Business, and so on), Global Village Services, Grid Computing, IBM SSME, Business Modelling, Agile Enterprise Architecting through ISRUP, and Globalization Governance through IT/IS Services.
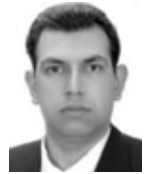

**Y. Rastegari** is PhD. Candidate of Computer Engineering Department of Shahid Beheshti University (SBU) in Tehran, Iran. He is also member of Automated Software Engineering Research Group (ASER) and Information System Architecture Research Center at SBU. His research interests include Service Oriented Architecture, Software Architecture, Software Product Line and Ontology Engineering. He is working on Service Choreography Adaptation challenge for Collaborative Business Processes.
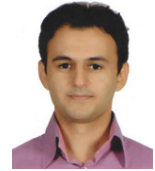

**E. Pejman** received his B.Sc. degree in Computer Science and Software Engineering from the Yazd University, Yazd, Iran in 2009 and his M.Sc. degree in Computer Science and Software Engineering from the Azad University-South Tehran Branch, Tehran, Iran in 2012. His research interests include web services, QoS-based service composition, robotics and image processing.
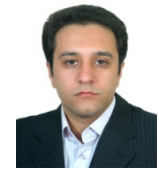